\documentclass[12pt]{article}
\usepackage{amssymb}
\usepackage{epsfig}
\usepackage{bbm}

\usepackage{graphicx}

\def\L{\mathcal L}
\def\e{\varepsilon}

\newcommand{\wt}{\widetilde}

\begin{document}

\def\a{\alpha}
\def\b{\beta}
\def\c{\chi}
\def\d{\delta}
\def\e{\epsilon}
\def\f{\phi}
\def\g{\gamma}
\def\h{\eta}
\def\i{\iota}
\def\j{\psi}
\def\k{\kappa}
\def\l{\lambda}
\def\m{\mu}
\def\n{\nu}
\def\o{\omega}
\def\p{\pi}
\def\q{\theta}
\def\r{\rho}
\def\s{\sigma}
\def\t{\tau}
\def\u{\upsilon}
\def\x{\xi}
\def\z{\zeta}
\def\D{\Delta}
\def\F{\Phi}
\def\G{\Gamma}
\def\J{\Psi}
\def\L{\Lambda}
\def\O{\Omega}
\def\P{\Pi}
\def\Q{\Theta}
\def\S{\Sigma}
\def\U{\Upsilon}
\def\X{\Xi}

%Varletters
\def\ve{\varepsilon}
\def\vf{\varphi}
\def\vr{\varrho}
\def\vs{\varsigma}
\def\vq{\vartheta}

\def\dg{\dagger}                                     % hermitian conjugate
\def\ddg{\ddagger}                                   % double dagger
\def\wt#1{\widetilde{#1}}                    % big tilde
\def\mt{\widetilde{m}_1}
\def\mti{\widetilde{m}_i}
\def\rt{\widetilde{r}_1}
\def\mtt{\widetilde{m}_2}
\def\mttt{\widetilde{m}_3}
\def\rtt{\widetilde{r}_2}
\def\mb{\overline{m}}
\def\VEV#1{\left\langle #1\right\rangle}        % < >
\def\be{\begin{equation}}
\def\ee{\end{equation}}
\def\ds{\displaystyle}
\def\ra{\rightarrow}

\def\bea{\begin{eqnarray}}
\def\eea{\end{eqnarray}}
\def\NO{\nonumber}
\def\Bar#1{\overline{#1}}

% Journal abbreviations (preprints)

\def\pl#1#2#3{Phys.~Lett.~{\bf B {#1}} ({#2}) #3}
\def\np#1#2#3{Nucl.~Phys.~{\bf B {#1}} ({#2}) #3}
\def\prl#1#2#3{Phys.~Rev.~Lett.~{\bf #1} ({#2}) #3}
\def\pr#1#2#3{Phys.~Rev.~{\bf D {#1}} ({#2}) #3}
\def\zp#1#2#3{Z.~Phys.~{\bf C {#1}} ({#2}) #3}
\def\cqg#1#2#3{Class.~and Quantum Grav.~{\bf {#1}} ({#2}) #3}
\def\cmp#1#2#3{Commun.~Math.~Phys.~{\bf {#1}} ({#2}) #3}
\def\jmp#1#2#3{J.~Math.~Phys.~{\bf {#1}} ({#2}) #3}
\def\ap#1#2#3{Ann.~of Phys.~{\bf {#1}} ({#2}) #3}
\def\prep#1#2#3{Phys.~Rep.~{\bf {#1}C} ({#2}) #3}
\def\ptp#1#2#3{Progr.~Theor.~Phys.~{\bf {#1}} ({#2}) #3}
\def\ijmp#1#2#3{Int.~J.~Mod.~Phys.~{\bf A {#1}} ({#2}) #3}
\def\mpl#1#2#3{Mod.~Phys.~Lett.~{\bf A {#1}} ({#2}) #3}
\def\nc#1#2#3{Nuovo Cim.~{\bf {#1}} ({#2}) #3}
\def\ibid#1#2#3{{\it ibid.}~{\bf {#1}} ({#2}) #3}

\title{
{\normalsize \mbox{ }\hfill
\begin{minipage}{5cm}
CERN-PH-TH/2008--193
\end{minipage}}\\
\vspace*{10mm}
\bf Successful type I Leptogenesis \\
with $SO(10)$-inspired mass relations}
\author{
{\Large Pasquale Di Bari$^a$ and Antonio Riotto$^{a,b}$}
\\
$^a$
{\it INFN, Sezione di Padova},
{\it Dipartimento di Fisica Galileo Galilei} \\
{\it  Via Marzolo 8, I-35131 Padua, Italy}
 \\
$^b$
{\it CERN, PH-TH Division,  CH-1211, Geneva 23, Switzerland}
}

\maketitle \thispagestyle{empty}

\vspace{-5mm}
%\centerline{\date{\today}}

\begin{abstract}
\noindent
It is well-known that thermal leptogenesis through the decays of the lightest
right-handed neutrinos encounters serious difficulties when
$SO(10)$-inspired mass conditions are imposed on the Dirac neutrino
mass matrix and light neutrino masses are generated through the type I see-saw mechanism.
We show that these can be circumvented when the production from the next-to-lightest
right-handed neutrinos and flavor effects are properly taken into account.
Some conditions on the low energy parameters have to be satisfied in order for
inverse processes involving the lightest right-handed  neutrino not to wash-out the asymmetry.
In particular we find $m_1\gtrsim 10^{-3}\,{\rm eV}$, where $m_1$ is the mass of the
lightest left-handed neutrino and that  non-vanishing values of the mixing angle $\theta_{13}$
are preferred in the case of a normal fully hierarchical spectrum of light neutrinos.
\end{abstract}

\newpage

%%%%%%%%%%%%%%%%%%%%%%%%%%%%%%%%%%%%%%%%%%%%%%%%%%
\section{Introduction}
%%%%%%%%%%%%%%%%%%%%%%%%%%%%%%%%%%%%%%%%%%%%%%%%%%

Thermal Leptogenesis \cite{FY,review} is an elegant model  to explain the observed
baryon asymmetry of the Universe and a direct consequence
of the see-saw mechanism \cite{seesaw} for the explanation of the neutrino masses.
Current data on neutrino masses are not only compatible with the
minimal version of leptogenesis,  but even exhibit interesting correlations
that support the picture. The strongest one is perhaps that the solar and the atmospheric
neutrino mass scales are one-two order of magnitude larger than the so called equilibrium
neutrino mass scale setting the effectiveness of the wash-out processes. In this
way successful leptogenesis is possible and at the same time its predictions
become particularly simple and, more importantly, do not depend  on the initial conditions.

On the other hand there are  some well known drawbacks
in the minimal version of thermal leptogenesis
that relies on the simplest version of the see-saw mechanism, the type I.
The most serious one is a potential conflict between leptogenesis and
$SO(10)$ grand-unified theories \cite{orlof,branco}, commonly regarded as the most
attractive way to embed the see-saw mechanism. Indeed, in a traditional version of leptogenesis, where
the spectrum of right-handed (RH)  neutrinos is hierarchical and the asymmetry is produced
from the decays of the lightest ones, there is a stringent lower bound on their mass \cite{di},
$M_1 > {\cal O}(10^9)\,{\rm GeV}$, for a sufficiently large baryon asymmetry to be produced.
On the other hand, $SO(10)$ grand-unified theories  typically yield, in their simplest version
and for the measured values of the neutrino mixing parameters,
a hierarchical spectrum with the RH neutrino masses proportional
to the squared of the up-quark masses, leading to
$M_1={\cal O}(10^5)\,{\rm GeV}$ and
to a final asymmetry that falls a few orders of magnitude below the observed one.

Of course,  for very fine tuned choices of the parameters,
it is possible to get a degenerate RH neutrino spectrum that produce an enhancement
of the $C\!P$ asymmetries and consequently of the final asymmetry,
see Ref.  \cite{afs} for a detailed analysis. Furthermore,
non-minimal versions of leptogenesis based on a mixed type I plus type II
see-saw mechanism can be adopted \cite{hosteins,proceedings,french}.
It is nevertheless   fair to say that a traditional type I picture of leptogenesis,
where the spectrum is hierarchical and the asymmetry is generated from the decays of the
lightest RH neutrinos, encounters serious difficulties within grand-unified $SO(10)$ theories.

Two crucial aspects have been  usually neglected in the previous studies though.
First, the contribution to the final baryon asymmetry
  from the decays of the next-to-lightest
  RH neutrinos $N_2$ \cite{geometry} and, secondly, the effect of
  flavor in thermal leptogenesis \cite{flavour,flavour1}.
  In the case in which the baryon asymmetry is generated by the $N_2$'s,
  the above-mentioned  lower bound on $M_1$ disappears and is replaced by a lower bound on $M_2$,
  the mass of the next-to-lightest RH neutrino. In the one-flavor approximation,
  that is when flavor effects are not taken into account, one can show that the baryon
  asymmetry generated by the $N_2$ decays is generically  maximized for models
  characterized by having  $m_1\propto M_1^{-1}$, where $m_1$ is the lightest left-handed (LH)
  neutrino mass eigenvalue. Unfortunately, in their simplest version,
$SO(10)$ models yield different neutrino mass relations, where $m_1\propto M_3^{-1}$,
  with $M_3$ the mass of the heaviest RH neutrino.
  In such a case, the baryon asymmetry produced by the $N_2$'s is
subsequently washed-out by the interactions involving the lightest RH neutrinos.
  This leads to the conclusion  that $N_2$ thermal leptogenesis
  in the one-flavor approximation
  and with $SO(10)$ conditions is not able to explain the observed asymmetry.
 Flavor effects may help in this regard. Indeed, one can imagine the situation where the
 next-to-lightest RH neutrino decays predominantly in one particular flavor and that
 the wash-out mediated by the lightest RH neutrinos is inefficient along that flavor.
 This point was first made  in Ref.  \cite{vives}
 for generic models with a strong hierarchy in the RH Majorana masses arising
 when the up-quark--neutrino Yukawa unification is imposed.
 The importance of $N_2$-induced leptogenesis was  also  stressed in \cite{nardiN2}.

  The question we would like to answer in this paper is whether  thermal leptogenesis
  mediated by the next-to-lightest RH neutrinos with flavor effects
  can effectively work within  $SO(10)$ grand-unified models where
  light LH neutrino masses are generated via the type I see-saw
 mechanism.   In Ref. \cite{vives} there were no
  specific quantitative and full calculations performed within $SO(10)$.  It was only shown that in general
   the wash-out from the lightest RH neutrinos may be small in some flavor even though
   the total wash-out is strong, while the
$C\!P$ asymmetry generated by the decays of the next-to-lightest RH neutrino was assumed to be maximal.
The problem, however, requires more care. Indeed, successful leptogenesis requires
three conditions: first, the wash-out from the lightest RH neutrinos needs to be  tiny in some flavor
    for some choice of the parameters; secondly,
a large enough $C\!P$ asymmetry in that flavor must be
 produced from the decays of the next-to-lightest
RH neutrinos and, third, the  flavor asymmetry  should  not be
washed out by the interactions mediated
by the next-to-lightest RH neutrinos themselves.
    It is the interplay of all these three requirements
which determines whether successful leptogenesis may be achieved.

   In this Letter we show that there are regions in the parameter space where type I thermal
   leptogenesis with hierarchical RH neutrino masses is successful imposing simple
   $SO(10)$-inspired conditions on the neutrino Dirac mass matrix.
   Furthermore, we will show that requiring successful leptogenesis
     leads to  interesting predictions on the low energy neutrino parameters
    that will be tested in future experiments. A more quantitative detailed
    analysis involving the many low-energy parameters is in progress \cite{inprep}.

    The paper is organized as follows. In section 2 we describe the see-saw mechanism when
    $SO(10)$-like conditions are imposed on the Dirac mass matrix.
    In section 3 we describe the generic features of $N_2$-leptogenesis
    with such $SO(10)$-inspired mass relations.
    Finally, in section 4 we present our conclusions.

%%%%%%%%%%%%%%%%%%%%%%%%%%%%%%%%%%%%%%%%%%%%%%%%%%%%%%%%%%%%%%%%%
\section{See-saw mechanism with hierarchical Dirac mass spectrum}
%%%%%%%%%%%%%%%%%%%%%%%%%%%%%%%%%%%%%%%%%%%%%%%%%%%%%%%%%%%%%%%%%

The see-saw mechanism is based on a simple extension of the Standard Model
where three RH neutrinos $N_{i}$, $i=1,2,3$ (as nicely predicted
within $SO(10)$), with a Majorana mass mass matrix $M$ and Yukawa
couplings $h$ to leptons and Higgs are added.
% 3 da SO (1O)
After spontaneous symmetry breaking,
a Dirac mass term $m_D=h\, v$, is generated
by the vacuum expectation value (VEV)  $v=174$ GeV of the Higgs boson.
In the see-saw limit, $M\gg m_D$, the spectrum of neutrino mass eigenstates
splits in two sets: three very heavy neutrinos, $N_1,N_2$ and $N_3$
respectively with masses $M_1\leq M_2 \leq M_3$ almost coinciding with
the eigenvalues of $M$, and three light neutrinos with masses $m_1\leq m_2\leq m_3$,
the eigenvalues of the light neutrino mass matrix
given by the see-saw formula \cite{seesaw},
\be
m_{\nu}= - m_D\,{1\over M}\,m_D^T \, .
\ee
A parametrization of the Dirac mass matrix is
obtained in terms of the orthogonal matrix $\Omega$ \cite{casas}
\be\label{h}
m_D=U\,D_m^{1/2}\,\O\, D_M^{1/2} \,  ,
\ee
where we defined
$D_m\equiv {\rm diag}(m_1,m_2,m_3)$ and $D_M\equiv {\rm diag}(M_1,M_2,M_3)$.
Neutrino oscillation experiments measure two
neutrino mass-squared differences. For normal schemes one has
$m^{\,2}_3-m_2^{\,2}=\Delta m^2_{\rm atm}$ and
$m^{\,2}_2-m_1^{\,2}=\Delta m^2_{\rm sol}$,
whereas for inverted schemes one has
$m^{\,2}_3-m_2^{\,2}=\Delta m^2_{\rm sol}$
and $m^{\,2}_2-m_1^{\,2}=\Delta m^2_{\rm atm}$.
For $m_1\gg m_{\rm atm} \equiv
\sqrt{\Delta m^2_{\rm atm}+\Delta m^2_{\rm sol}}=
(0.050\pm 0.001)\,{\rm eV}$ \cite{oscillations}
the spectrum is quasi-degenerate, while for
$m_1\ll m_{\rm sol}\equiv \sqrt{\D m^2_{\rm sol}}
=(0.00875\pm 0.00012)\,{\rm eV}$ \cite{oscillations}
it is fully hierarchical (normal or inverted).
Here we will restrict ourselves to the case of normal schemes.
The most stringent upper bound on the
absolute neutrino mass scale comes from
cosmological observations.
Recently, a conservative upper bound on the sum of neutrino masses,
$\sum_i\,m_i\leq 0.61\,{\rm eV}\; (95\%\, {\rm CL})$,
has been obtained by the WMAP collaboration combining CMB,
baryon acoustic oscillations and supernovae type Ia observations \cite{WMAP5}.
Considering that it falls in the quasi-degenerate regime, it straightforwardly translates into
\be\label{upperbound}
m_1 < 0.2\,{\rm eV} \;\; (95\%\, {\rm CL}) \, .
\ee
The matrix $U$ diagonalizes the light neutrino mass matrix $m_{\nu}$, so that
\begin{equation}
U^{\dagger}\,m_{\nu}\,U^{\star}=-D_m
\end{equation}
and it can be identified with the lepton mixing matrix in a basis where
the charged lepton mass matrix is diagonal.
We will adopt the following parametrization  for the matrix $U$
in terms of the mixing angles, the Dirac phase $\delta$ and
 the Majorana phases $\rho$ and $\sigma$ \cite{PDG}
\begin{equation}\label{Umatrix}
U=\left( \begin{array}{ccc}
c_{12}\,c_{13} & s_{12}\,c_{13} & s_{13}\,e^{-{\rm i}\,\d} \\
-s_{12}\,c_{23}-c_{12}\,s_{23}\,s_{13}\,e^{{\rm i}\,\d} &
c_{12}\,c_{23}-s_{12}\,s_{23}\,s_{13}\,e^{{\rm i}\,\d} & s_{23}\,c_{13} \\
s_{12}\,s_{23}-c_{12}\,c_{23}\,s_{13}\,e^{{\rm i}\,\d}
& -c_{12}\,s_{23}-s_{12}\,c_{23}\,s_{13}\,e^{{\rm i}\,\d}  &
c_{23}\,c_{13}
\end{array}\right)
\cdot {\rm diag}\left(e^{i\,\rho}, 1, e^{i\,\sigma}
\right)\, ,
\end{equation}
where $s_{ij}\equiv \sin\theta_{ij}$,
$c_{ij}\equiv\cos\theta_{ij}$ and we will adopt the
following 2-$\sigma$ ranges compatible with the results
from neutrino oscillation experiments \cite{oscillations}:
\be\label{twosigma}
\theta_{12}= (31.3^\circ-36.3^\circ)  \, , \;\;\;
\theta_{23}= (38.5^\circ-52.5^\circ) \, , \;\;\;
\theta_{13}= (0^\circ-11.5^\circ) \, .
\ee
With the adopted convention for the light neutrino masses, $m_1<m_2<m_3$,
this parametrization is valid only for normal hierarchy, while for inverted
hierarchy one has to perform a column cyclic permutation.

Within leptogenesis, the see-saw mechanism is also able to explain
the observed baryon asymmetry of the Universe
\cite{WMAP5}
\be\label{etaBobs}
\eta_B^{\rm CMB} = (6.2 \pm 0.15)\times 10^{-10} \, .
\ee
The predicted baryon-to-photon ratio $\eta_B$ is
related to the final value of the final $(B-L)$ asymmetry $N^{\rm f}_{B-L}$
by the relation \cite{review}
\be\label{etaB}
\eta_B \simeq 0.96\times 10^{-2} N_{B-L}^{\rm f}
\ee
and we will impose a 2-$\sigma$ bound $\eta_B>5.9\times 10^{-10}\,{\rm GeV}$
for successful leptogenesis. Notice that the small experimental error on
$m_{\rm atm}$ can be neglected in our case and we use everywhere
$m_{\rm atm}=0.05\,{\rm eV}$.

In general the final asymmetry in leptogenesis depends on all parameters
in the right-hand side of (\ref{h}),  in particular on the `high-energy' parameters
in the orthogonal matrix $\O$ and on the three RH neutrino masses $M_i$. In this way imposing
successful leptogenesis does not yield definite predictions on the low energy parameters.

However some theoretical input in $m_D$ can produce relations
able to express the nine high energy parameters, six in $\O$ plus the three $M_i$,
through the low energy parameters and when 
successful leptogenesis is required, interesting predictions
on the low energy parameters can follow.

This is exactly what happens within $SO(10)$ models, where the theoretical
input at the grand-unified scale is conveniently plugged into $m_D$. In this sense, it
proves useful to adopt the bi-unitary parametrization
\be
m_D=V_L^{\dagger}\,D_{m_D}\,U_R \, ,
\ee
where $V_L$ and $U_R$ are two unitary matrices that diagonalize
$m_D$ and $D_{m_D}$  is the diagonal matrix whose elements are the
eigenvalues of $m_D$: $D_{m_D}\equiv {\rm diag}(\l_{D1},\l_{D2},\l_{D3})$.

The matrix $V_L$ is analogous to the CKM matrix in the quark sector.
The important point is that, once the LH neutrino mass matrix $m_\nu$
is chosen, and for a given $V_L$ and $D_{m_D}$, the masses of the heavy RH neutrinos are fixed together with the matrix $U_R$.
Indeed, the see-saw relation can be always written in a basis where
$M$ is diagonal. Moreover, in the basis where the charged lepton mass
matrix is diagonal, the neutrino mass matrix $m_{\nu}$ is made diagonal
by the lepton mixing matrix $U$, $m_{\nu}=-U\,D_m\,U^T$.
In this way it is easy to see that the matrix
\be
M^{-1} \equiv D^{-1}_{m_D}\,V_L\,U\,D_m\,U^T\,V_L^T\,D^{-1}_{m_D}
\ee
is diagonalized by $U_R$ and the eigenvalues are the $M_i^{-1}$,
\be
M^{-1}=U_R\,D_M^{-1}\,U_R^T \, .
\ee
Notice that  we are defining the $M_i$'s to be real and positive while
we incorporate the corresponding phases into the matrix $U_R$.

Given the theoretical inputs from $SO(10)$, indicating $V_L$ and $D_{m_D}$, and upon
diagonalizing $M^{-1}$, one obtains the matrix $U_R$ and the RH neutrino
masses $M_i$ as a function of the low energy parameters, the light neutrino masses
$m_i$ and the mixing parameters in $U$. From the relation (\ref{h}),
one can then also calculate the orthogonal matrix $\O$ as a function of the
low energy parameters,
\be
\O=D_m^{-{1\over 2}}\,U^{\dagger}\,V_L^{\dagger}\,D_{m_D}\,U_R\,D_M^{-{1\over 2}} \, .
\ee
Now, in  minimal $SO(10)$ scenarios \cite{goran},
it is expected that a small misalignment between the charged
lepton mass matrix and the Dirac neutrino mass matrix $m_D$,
similar to that in the quark sector. Under
the assumption of small misalignment,
$V_L$ should be close to the unity matrix and we will adopt this
assumption in the following. We have explicitly checked that small departures
from this assumption, {\it e.g.} $V_L=V_{\rm CKM}$,
do not alter our conclusions \cite{inprep}.
As for the mass eigenvalues, since  fermion families are 16-dimensional
spinors of $SO(10)$, from $16_F\times 16_F=10_H+120_H+126_H$,
one deduces that the Yukawa coupling matrices may get several
contributions. If the dominant one comes from the 10-dimensional Higgs multiplet,
then the relation between the up-quark mass matrix $m_u$ and the
Dirac mass matrix $m_D$ reads $m_u=m_D$. On the other hand,
if fermion masses are generated by the VEV of a 126-dimensional Higgs,
then the $SU(4)$ symmetry relation holds $3 m_u=-m_D$.
A sort of mixture between these two possibilities is achieved if
the 120-dimensional Higgs dominates the contribution to $m_D$.
Of course, the total dominance of one Higgs representation
is excluded because this would mean that all the fermion mass
matrices would be simultaneously diagonalized. Besides bad mass relations, this would imply
no quark and no lepton mixings in the weak currents.
The fact that the minimal theory must have (at least) two Higgs
representations and the fact that one can also envisage the contribution to
$m_D$ from non-renormalizable interactions, lead us to assume that the Yukawa couplings are
only approximately the same for the up-type quarks and neutrinos.
We will therefore just assume a hierarchical pattern with
$\l_{D3}\gg \l_{D2} \gg \l_{D1}$. This holds using a parametrization
\be\label{SO(10)}
\l_{D1}= \alpha_1\,m_u , \;\; \l_{D2}= \alpha_2\, m_c ,\;\; \l_{D3}= \alpha_3\,m_t \, ,
\ee
where for the up-quark masses we use the reference values
$m_u=1\,{\rm MeV}$, $m_c=400\,{\rm MeV}$ and $m_t={\rm 100 GeV}$,
approximately coinciding with the up-type quark masses at $T\sim 10^9\,{\rm GeV}$ \cite{fusaoka},
and where the coefficients $\alpha_{i}={\cal O}(1)$.
With this assumption the RH neutrino mass spectrum is hierarchical and
of the form (for generic expressions in terms of the low energy parameters,
see Ref. \cite{afs})
\be
\label{alpha}
M_1\,:\,M_2\,:\,M_3=(\alpha_1\,m_u)^2\,:\,(\alpha_2\,m_c)^2\,:\,(\alpha_3\;m_t)^2\, .
\ee
In Fig.~1 we show the dependence of the three $M_i/\alpha_i^2$ on $m_1$
for two different sets of values of ${\theta_{23},\theta_{12},\theta_{13},\delta,\rho,\sigma}$,
as indicated in the figure caption.
\begin{figure}
\psfig{file=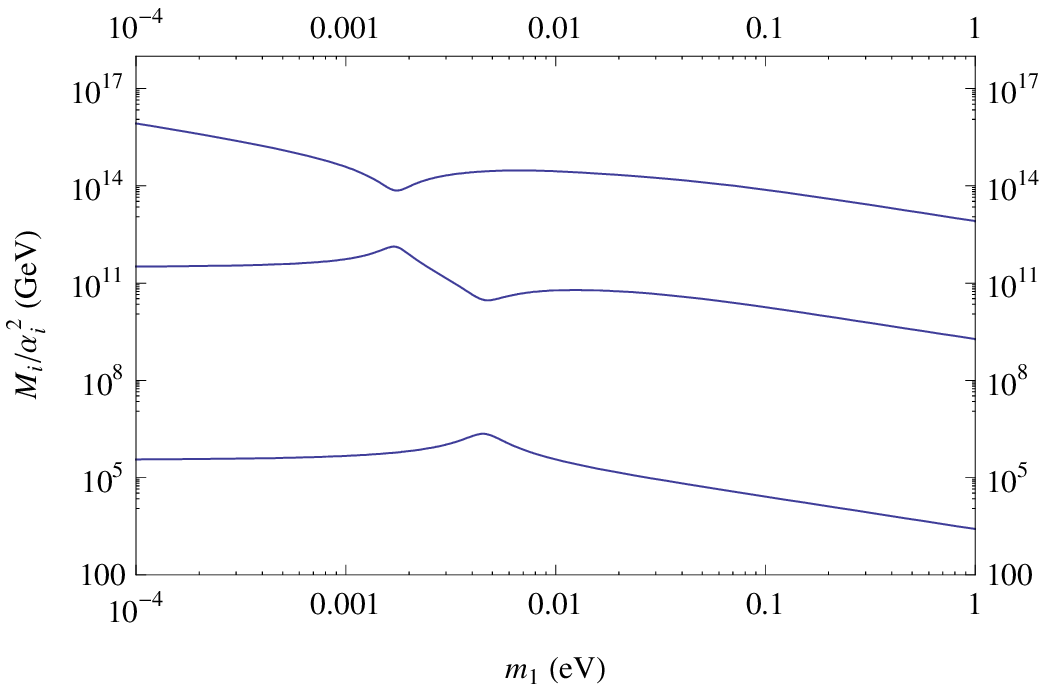,height=42mm,width=50mm}
\hspace{3mm}
\psfig{file=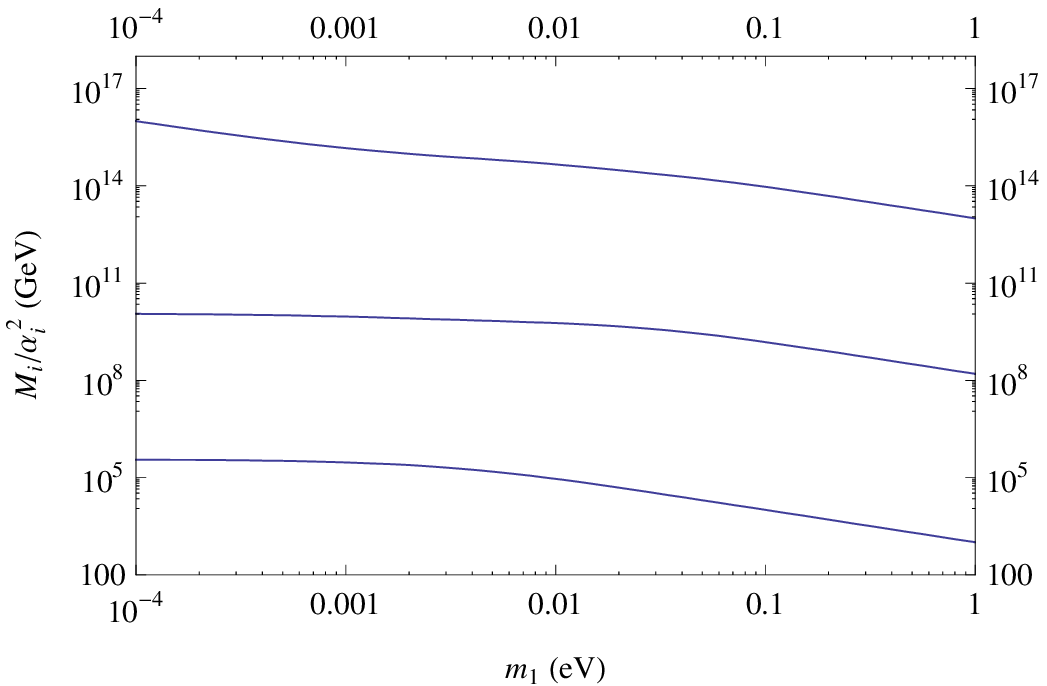,height=42mm,width=50mm}
\hspace{3mm}
\psfig{file=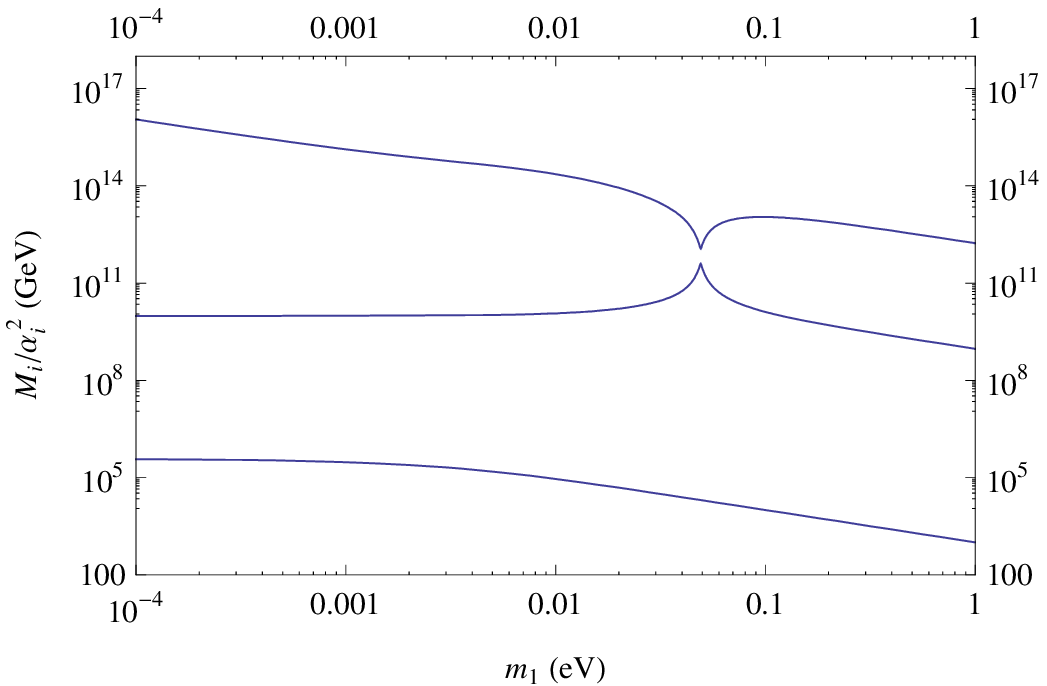,height=42mm,width=50mm}
\caption{Dependence of the three $M_i/\alpha_i^2$ on $m_1$ for $\theta_{13}=5^{\circ},\theta_{23}=40^{\circ}$
and $\theta_{12}=33.5^{\circ}$ and for three different choices of the phases:
$\delta=\sigma=0^{\circ}, \rho=1.5^{\circ}$ (left); $\delta=5.86^{\circ},
\rho=\sigma=3^{\circ}$ (center); $\delta=\pi/3,\rho=0.02^{\circ},\sigma=\pi/2$ (right).}
\label{fig:Mi}
\end{figure}
One can see how non-vanishing values of the phases can enhance
the values of the RH neutrino masses for particular values of $m_1$.

We stress that  for our considerations, the values of $\alpha_1$ and $\alpha_3$
are irrelevant (unless $\alpha_1$ is unrealistically large to push $M_1$ from $\sim 10^5\,{\rm GeV}$
above the lower bound $\sim  10^{9}$ GeV to achieve successful $N_1$ leptogenesis).
On the other hand, the value of $\alpha_2$ is only relevant
to set the scale of the mass $M_2\simeq 2(\alpha_2\,m_c)^2/m_3$ (valid for $\theta_{13}\simeq 0$)
of the next-to-lightest RH neutrino mass, but it does not alter
other quantities crucial for thermal leptogenesis,
such as the amount of wash-out from the lightest RH neutrinos.
Furthermore, as we will show, realistic values of $\alpha_2\gtrsim 3$ are already enough to
provide a sufficiently large baryon asymmetry.
We now have all the ingredients to calculate
the final asymmetry within flavored $N_2$-leptogenesis.

%%%%%%%%%%%%%%%%%%%%%%%%%%%%%%%%%%%%%%%%%%%%%%%%%%%%%%%%%%%%%%%%%
\section{Thermal leptogenesis from next-to-lightest RH neutrinos}
%%%%%%%%%%%%%%%%%%%%%%%%%%%%%%%%%%%%%%%%%%%%%%%%%%%%%%%%%%%%%%%%%

Our working assumption is that the final asymmetry is dominantly produced from $N_2$-decays.
Having imposed the $SO(10)$ relations (\ref{SO(10)}),
the RH neutrino mass spectrum satisfies, except for special cases \cite{afs},
the condition  $M_1\ll 10^9\,{\rm GeV}\lesssim M_2 \lesssim 10^{12}\,{\rm GeV}$
and therefore, as explained above,   the asymmetry produced from $N_1$-decays
cannot reproduce the observed one.

As we outlined in the introduction, type I thermal leptogenesis in $SO(10)$
may be successful through the following chain of processes. The out-of-equilibrium decays
of the $N_2$'s produce at temperatures $T\sim M_2$  asymmetries in the two flavor regime,
therefore in the tauon flavor and into an asymmetry stored in leptons
that are a coherent over-position of electron and muon components.
Below $T\sim 10^9\,{\rm GeV}$ the three flavor regime holds and one has
distinct asymmetries in all three flavors.  Subsequently,
the interactions mediated by the $N_1$'s wash-out part of these asymmetries.
Successful leptogenesis may be achieved if
a sufficiently large asymmetry has been generated in a given
flavor from $N_2$ decays which is not washed-out by the $N_1$-mediated processes.

Since the asymmetry production from $N_2$ decays at $T\sim M_2$ occurs in a two-flavor
regime, when only the interactions mediated by the $\tau$-Yukawa couplings are in equilibrium,
the $(B-L)$ asymmetry from $N_2$-decays is  the sum of two contributions
\be\label{NBmLTM2}
N_{B-L}^{T\sim M_2} \simeq
\ve_{2\tau}\,\kappa(K_{2\tau})+ \ve_{2e+\mu}\,\kappa(K_{2e+\mu}) \, ,
\ee
where the flavored $C\!P$ asymmetries in the flavor $\alpha$ are defined as
\be
\ve_{2\a}\equiv -{\G_{2\alpha}-\overline{\G}_{2\alpha}
\over \G_{2}+\overline{\G}_{2}} \, .
\ee
Here $\G_{2\alpha}$ ($\overline{\G}_{2\alpha}$) is the partial decay rate of the $N_2$'s
into the $\alpha$-flavored lepton (anti-lepton) and $(\G_{2}+\overline{\G}_{2})$
is the total decay rate into lepton and anti-leptons. They
 can be calculated using the expression $(x_2=M_2^2/M_1^2)$ \cite{crv}
\begin{eqnarray}\label{eps2a}
\ve_{2\a}&\simeq&
\frac{3}{16 \p (h^{\dag}h)_{22}}  \left\{ {\rm Im}\left[h_{\a 2}^{\star}
h_{\a 3}(h^{\dag}h)_{2 3}\right] \frac{\x(x_3/x_2)}{\sqrt{x_3/x_2}}+
\frac{2}{3(x_3/x_2-1)}{\rm Im}
\left[h_{\a 2}^{\star}h_{\a 3}(h^{\dag}h)_{3 2}\right]\right\}\, , \nonumber\\
&& \\
\xi(x)&=&\frac{2}{3}x\left[(1+x)\ln\left(\frac{1+x}{x}\right)-\frac{2-x}{1-x}\right]\, .\nonumber
\end{eqnarray}
We are neglecting the term,  negligible for $M_2\gg M_1$,
arising from the interference  between the tree level contribution
with one loop graphs containing the lightest RH neutrino in the propagator.
Furthermore, in the two flavour regime, the asymmetry $\ve_{2e+\mu}$ stands for
$\ve_{2e+\mu}=\ve_{2e}+\ve_{2\mu}$.

In Fig.~2 we plotted the dependence of $\ve_2$, $\ve_{2\t}$ and $\ve_{2 e+\mu}$ on $m_1$
for the same values of the mixing angles as in Fig.~1. In the left (right) panel
the values of the phases are the same as in the center (right) panel of Fig.~1.
\begin{figure}
\begin{center}
\psfig{file=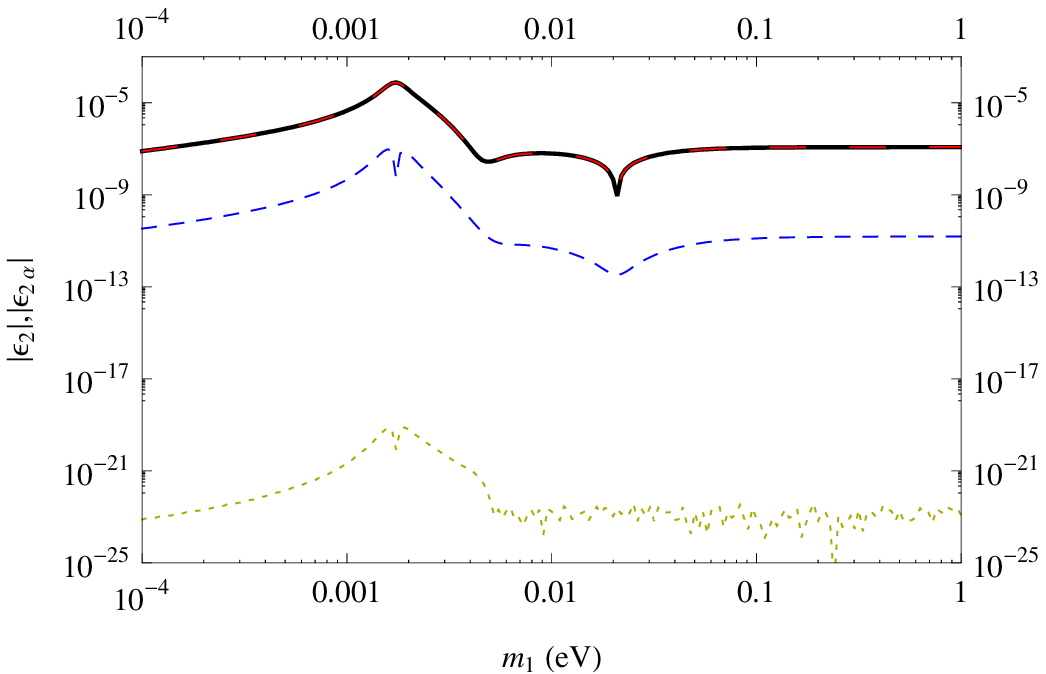,height=42mm,width=49mm}
\hspace{3mm}
\psfig{file=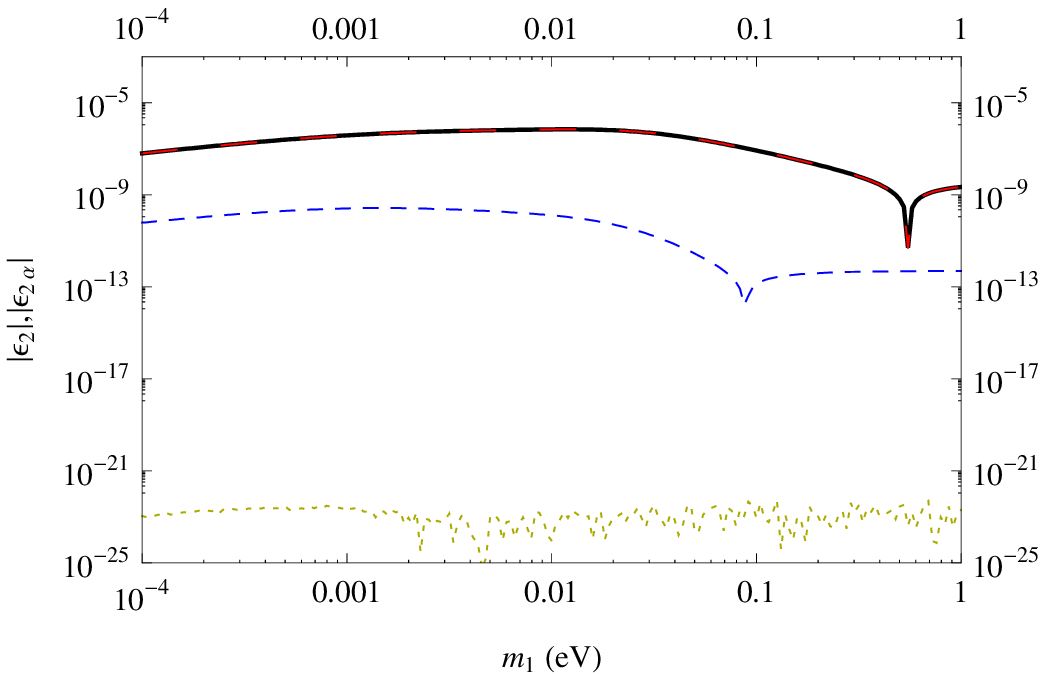,height=42mm,width=49mm}
\hspace{3mm}
\psfig{file=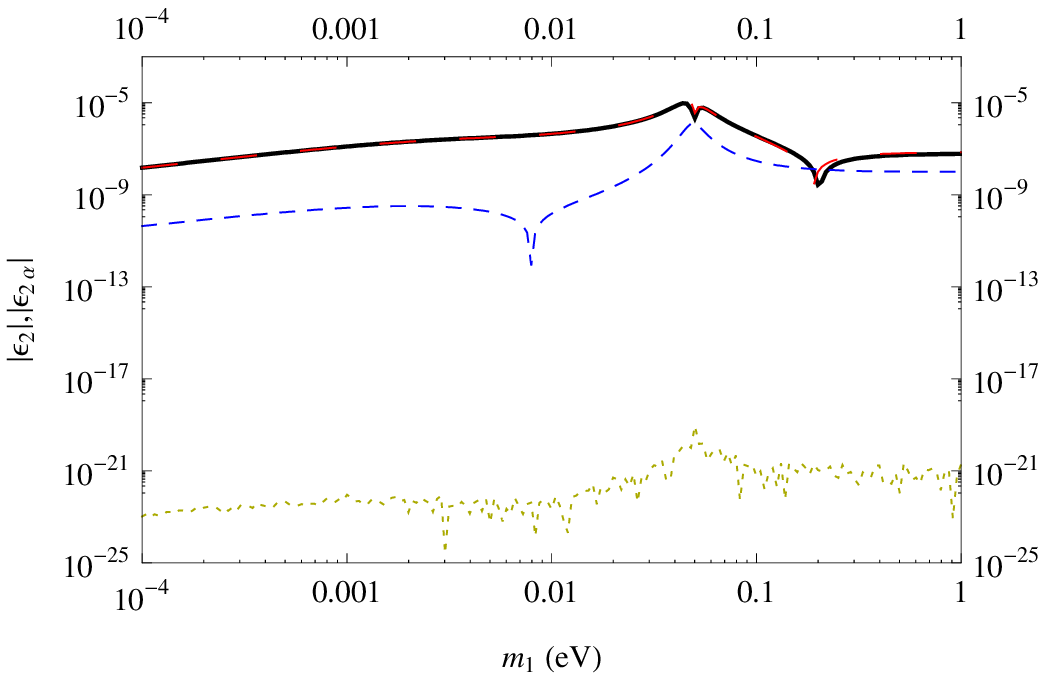,height=42mm,width=49mm}
\end{center}
\caption{Dependence of $|\ve_2|$ (black solid line), $|\ve_{2\tau}|$ (red long-dashed line),
$|\ve_{2\mu}|$ (blue short-dashed line) and $|\ve_{2e}|$ (yellow dotted line) on $m_1$
for the same values of the mixing angles and phases as in Fig.~1.
Moreover,  we have set $\alpha_i=({1,5,1})$.}
\label{fig:eps2}
\end{figure}
The efficiency factor is given by
\begin{eqnarray}
\k(x)&=&\frac{2}{x z_{\rm B}(x)}\left[1-{\rm exp}\left(-\frac{1}{2} x\, z_{\rm B}(x)\right)\right]\,,\nonumber\\
z_{\rm B}(x)&\simeq& 2+4 \,x^{0.13} \,{\rm exp}\left(-\frac{2.5}{x}\right)
\end{eqnarray}
and the flavored decay parameters are
\be
K_{i \alpha} = \left.\frac{\Gamma_{i\alpha}}{H}\right|_{T=M_2}=
                      \frac{\left|(m_D)_{\alpha i}\right|^2}{m_\star\, M_i}\, .
\ee
Here $H$ is the Hubble rate and
\be
m_\star=\frac{16\,\pi^{5/2}\sqrt{g_*}}{3\sqrt{5}}\frac{v^2}{M_{\rm Pl}}
\simeq 1.08\times 10^{-3}\,{\rm eV} \, ,
\ee
where $g_*$ is the effective relativistic degrees of freedom and $M_{\rm Pl}$ is the Planck mass.
Again, in the two-flavor regime, $K_{2e+\mu}$ stands for $K_{2e+\mu}=K_{2e}+K_{2\mu}$.
In Fig.~3 we plotted $K_2$, $K_{2\tau}$ and $K_{2e+\mu}$ versus $m_1$ for the
same values of the mixing angles and of the phases as in Fig.~2.
\begin{figure}
\begin{center}
\psfig{file=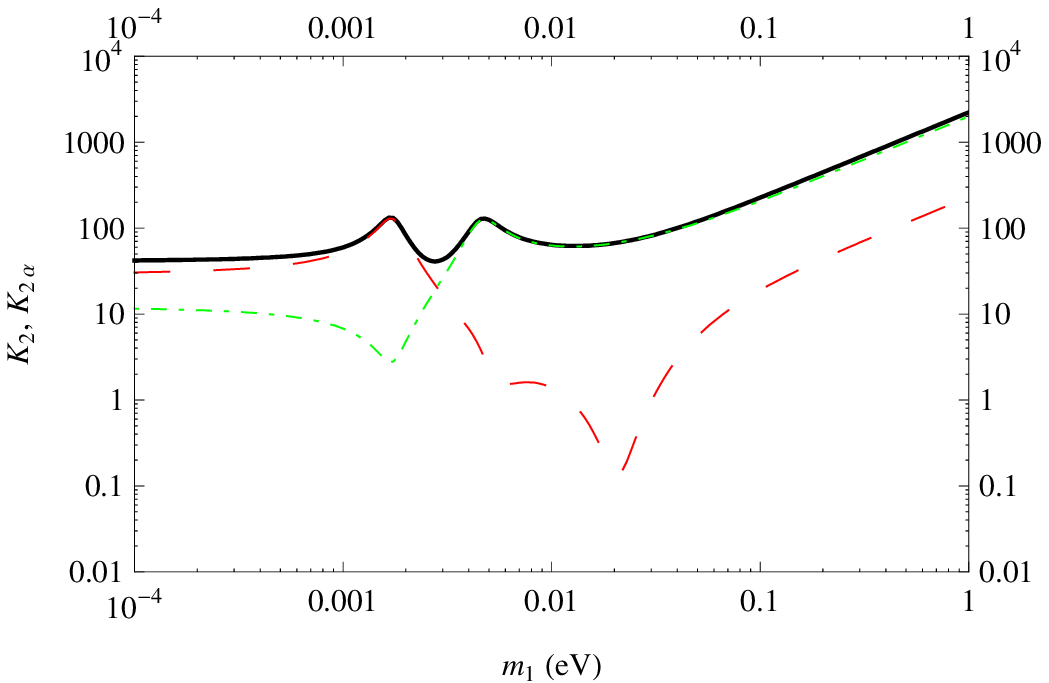,height=42mm,width=49mm}
\hspace{3mm}
\psfig{file=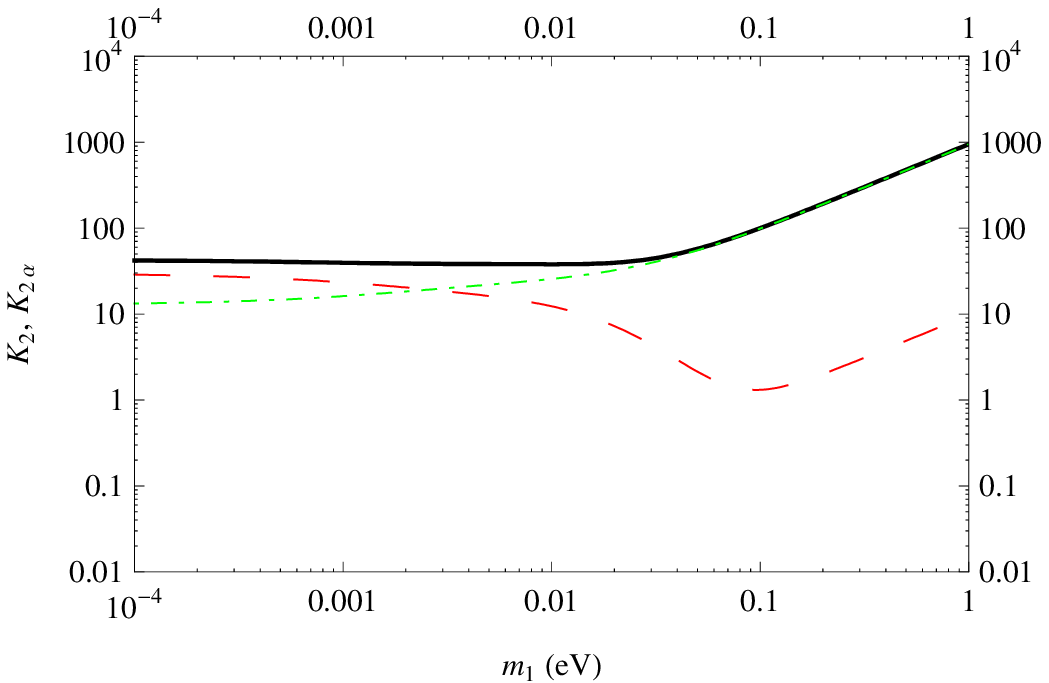,height=42mm,width=49mm}
\hspace{3mm}
\psfig{file=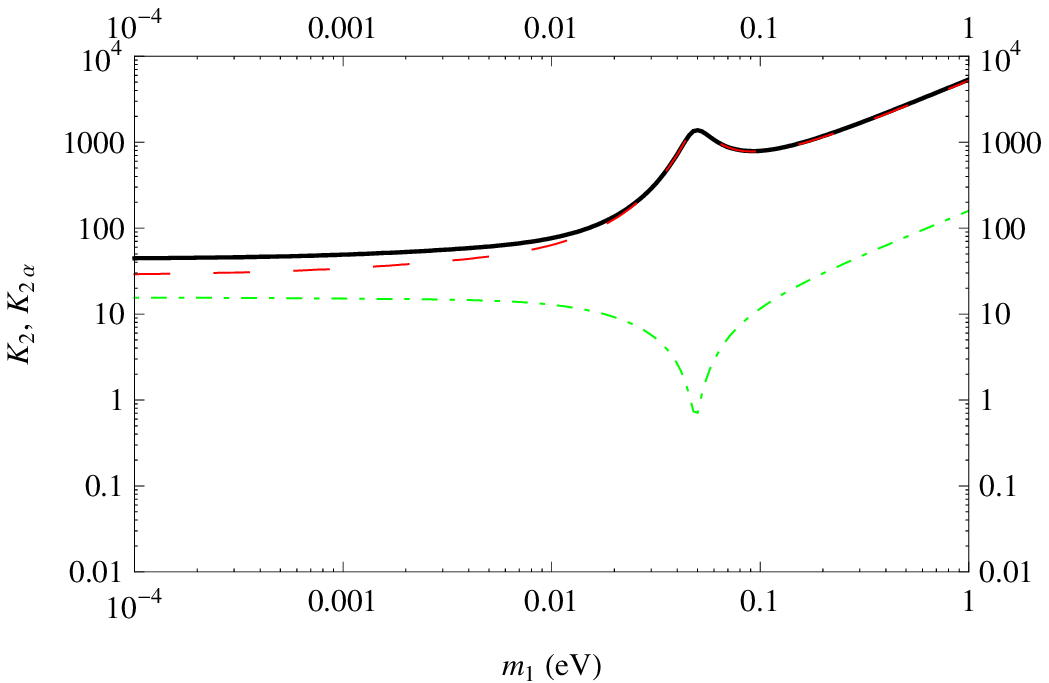,height=42mm,width=49mm}
\end{center}
\caption{Dependence of $K_2$ (black solid line), $K_{2\tau}$ (red long-dashed line) and
$K_{2e+\mu}$ (green dot-dashed line) on $m_1$ for the same three sets of values of the mixing
angles and phases as in previous figures.}
\label{fig:K2}
\end{figure}
It can be noticed how the wash-out from the same $N_2$ inverse processes is quite strong
in the $\tau$ flavor in the first (left panel) and in the third example (right panel).
In the second example (center panel) the wash-out is almost absent but there is
some $C\!P$ asymmetry suppression compared to the maximum value. A
suppression $\sim 1/30$ of $N_{B-L}^{T\sim M_2}$ is therefore actually unavoidable
compared to a maximum potential value. For this reason, and because $\ve_2 \propto M_2$,
the typical order of magnitude of $M_2$ for the mechanism to work is
 $M_2\sim 10^{11}\,{\rm GeV}$, two orders of magnitude higher than the usual
 lower bound in $N_1$-leptogenesis \cite{di} and one order of magnitude
 higher than the lower bound in $N_2$ leptogenesis without any condition
 on $m_D$ \cite{geometry}.
In Fig.~4 we finally plotted  $N_{B-L}^{T\sim M_2}$ (dashed line) for the same values of
the parameters as in Fig.~2 and in Fig.~3.
\begin{figure}
\begin{center}
\psfig{file=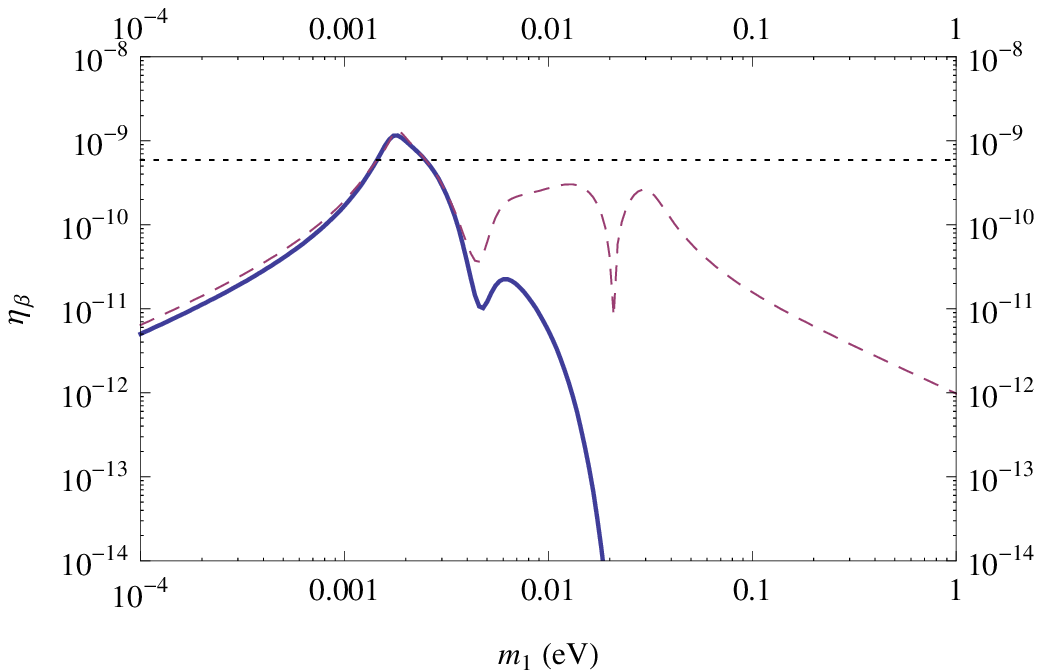,height=42mm,width=49mm}
\hspace{3mm}
\psfig{file=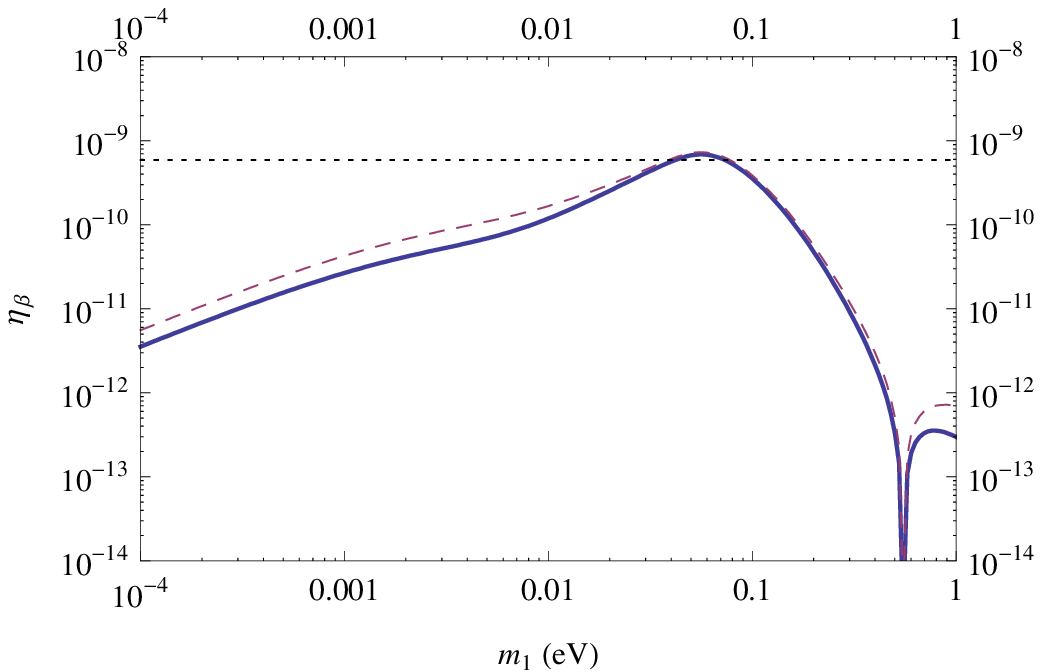,height=42mm,width=49mm}
\hspace{3mm}
\psfig{file=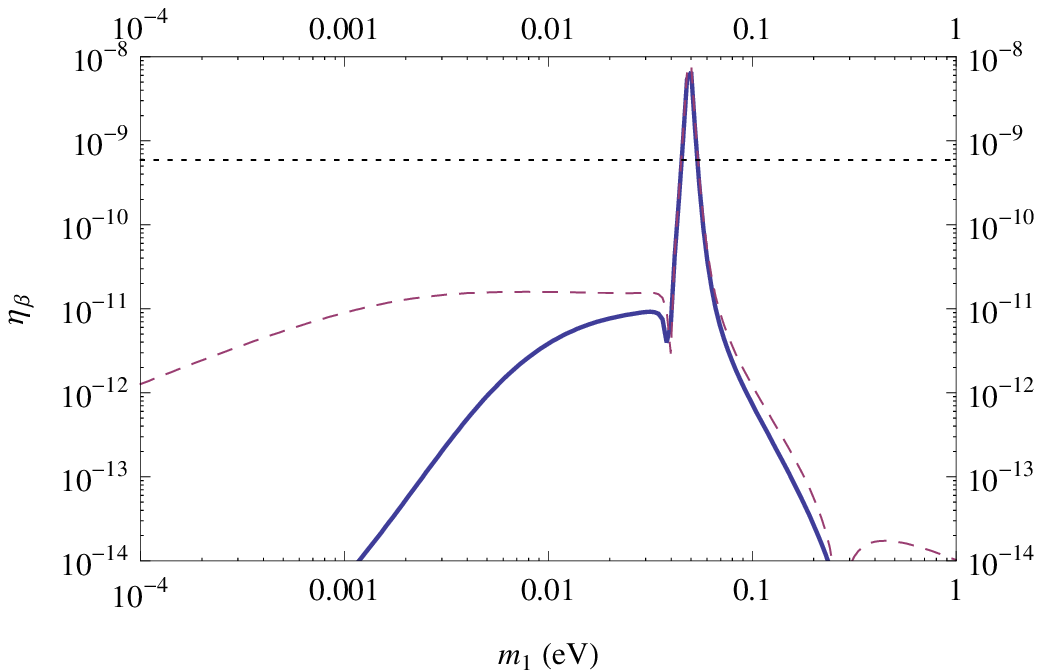,height=42mm,width=49mm}
\end{center}
\caption{Dependence of $\eta_B^{T\sim M_2}$
(dashed line) and of  $\eta_B^{\rm f}$  (solid line) on $m_1$ for the same
three sets of values of the parameters as in Fig.~2. The dotted line is the
$2\sigma$ lowest value $\eta_B^{CMB}=5.9\times 10^{-10}$ (cf. (\ref{etaBobs})).}
\label{fig:NBmL}
\end{figure}
We would like to stress that in the third example (right panel) for a very fine
tuned valued of $m_1$ and for the chosen values of $\a_i$, the values of $M_2$
and $M_3$ are equal and therefore there is a resonant $C\!P$ asymmetry
enhancement ($\xi(M_3^2/M_2^2)\gg 1$) just at the center of the peak in the right panel of Fig.~3.
However, far from the peak one has $M_3\gtrsim M_2$ implying no resonant
$C\!P$ asymmetry enhancement. Therefore it must be clearly stressed
that the success of the mechanism does not rely at all on a $C\!P$ asymmetry enhancement
due to $M_2\simeq M_3$. This will be even more clear when we will
perform a general scan in the space of parameters imposing a very restrictive
condition $M_3/M_2>10$ and we will find points with large enough asymmetry.

After the flavor asymmetries are generated at $T\sim M_2$, they are
subsequently washed out by the interactions mediated by the $N_1$'s.
This takes place at $T\sim M_1 \ll 10^9\,{\rm GeV}$, in the three flavor regime,
where the interactions
mediated by all the three charged lepton Yukawa interactions are in equilibrium.
Therefore, we calculate the final asymmetry by projecting
$N_{B-L}^{T\sim M_2}$ onto the three flavors using the expression
\be
N_{B-L}^{\rm f} \simeq
\ve_{2 e}\,\kappa(K_{2 e+\mu})\, e^{-{3\pi\over 8}\,K_{1 e}}+
\ve_{2\mu}\,\kappa(K_{2 e+\mu})\, e^{-{3\pi\over 8}\,K_{1 \mu}}+
\ve_{2 \tau}\,\kappa(K_{2 \tau})\,e^{-{3\pi\over 8}\,K_{1 \tau}} \, .
\ee
Notice that we are neglecting the matrix relating the
asymmetries stored in the lepton doublets to the $(B/3-L_{\alpha})$ asymmetries.
These would further decrease the wash-out from the lightest RH neutrino.
However, it has been noticed in \cite{aspects} that when the Higgs asymmetry
is taken into account as well, the diagonal terms sum up approximately to
unity and it is a good approximation to neglect their effect.

The first important thing to notice is that adopting  the
$SO(10)$ relation (\ref{SO(10)}) in the expression (\ref{eps2a})
one can see that\footnote{This hierarchy is lost
if the matrix $V_L$ differs significantly from the unity matrix.
To our understanding this is the case
in Ref. \cite{french} after translating their parametrization into ours.}
\be
\label{aa}
\ve_{2\t}:\ve_{2\mu}:\ve_{2e}=(\alpha_3\,m_{t})^2:(\alpha_2\,m_{c})^2:(\alpha_1\,m_{u})^2\, ,
\ee
from which we deduce that $\ve_{2\tau}\simeq \ve_2$ and that
most of the asymmetry generated by the $N_2$ decays is produced
predominantly  along the tau flavor. This can be clearly seen in the two
examples in Fig.~2. Furthermore,
in the limit of normal hierarchy for the light LH neutrinos,
adopting tri-bimaximal mixing for $U$,
($\theta_{13}=0, \theta_{12}\simeq 35.3^\circ,\theta_{23}=45^\circ$)
and setting, as we wrote earlier, the matrix $V_L$ equals to the unity matrix,  we find
\be
\ve_{2\t}\simeq\frac{9}{4\pi}\left(\frac{\alpha_2\, m_c}{v}\right)^2\frac{m_1}{m_3}\frac{{\rm Im}\left[\left(U_R^{\star}\right)^2_{32}\right]}{\left|
\left(U_R\right)_{32}\right|^2}\, .
\ee
Indeed, since $\ve_{2\t}\propto (M_2/M_3)$ and $M_3\propto m^{-1}_1$,
we immediately deduce that a large lepton asymmetry in the tau flavor may be
produced only for sufficiently large values of $m_1$. This is rather easy to understand.
If $m_1$ tends to zero, we go into the so-called decoupling limit, $M_2/M_3\simeq 0$.
As the $C\!P$ asymmetry needs (at least) two heavy states to be generated at the one-loop level,
and disregarding  the contribution from the $N_1$, $\ve_{2\t}$ must vanish.

The wash-out of such an  asymmetry by the next-to-lightest
neutrinos themselves is governed by the parameter
\begin{eqnarray}
K_{2\tau}&=&\frac{(\alpha_2\,m_c)^2}{m_\star}\frac{\left|\left(U_R\right)_{32}\right|^2}{M_2}\, ,\nonumber\\
|\left(U_R\right)_{32}| &\simeq&
\left| \frac{ \left(m_\nu\right)_{11} \left(m_\nu\right)_{23}- \left(m_\nu\right)_{12} \left(m_\nu\right)_{13}}{ \left(m_\nu\right)_{11} \left(m_\nu\right)_{22}- \left(\left(m_\nu\right)_{12}\right)^2}\frac{\alpha_2\, m_c}{\alpha_3\, m_t} \right| ,\nonumber\\
M_2&\simeq& \frac{(\alpha_2\, m_c)^2 \left|\left(m_\nu\right)_{11}\right|}{\left| \left(\left(m_\nu\right)_{12}\right)^2- \left(m_\nu\right)_{11} \left(m_\nu\right)_{22}\right|}\, .
\end{eqnarray}
For instance, using the same limits adopted
for the computation of the $C\!P$ asymmetry, we find
\be
K_{2\tau}\simeq\frac{m_3}{2m_\star\,\left[1-2s_{13}\cos(2\delta-\sigma)\right]}\, .
\ee
This gives $K_{2\tau}\simeq 25$, in agreement with the three numerical
examples in  Fig.~3. This
leads to a suppression in the efficiency factor given by
$\kappa(K_{2\tau})\lesssim 10^{-2}$ that is not enough
to prevent that the hierarchy in the $C\!P$ asymmetries translates into a hierarchy
in the flavor asymmetries. Therefore, the total asymmetry produced from $N_2$-decays at
$T\sim M_2/5$ is dominantly in the tau flavor. It follows that
in order for the mechanism to work we have to impose the subsequent
wash-out mediated by the $N_1$'s is inefficient along the tau flavor,
that is $K_{1\tau}\lesssim 1$. In Fig.~5 we plotted $K_1$ and the three $K_{1\alpha}$'s
depending on $m_1$ for the usual two sets of values. One can see that in these two examples
there are values for $m_1$ where $K_{1\tau}\lesssim 1$. Correspondingly
one can see in Fig.~4 (solid line) that $N_{B-L}^{\rm f}\simeq N_{B-L}^{T\sim M_2}$
meaning that the wash-out from the lightest RH neutrinos is circumvented:
the mechanism proposed in \cite{vives} is therefore  working in this case.
\begin{figure}
\begin{center}
\psfig{file=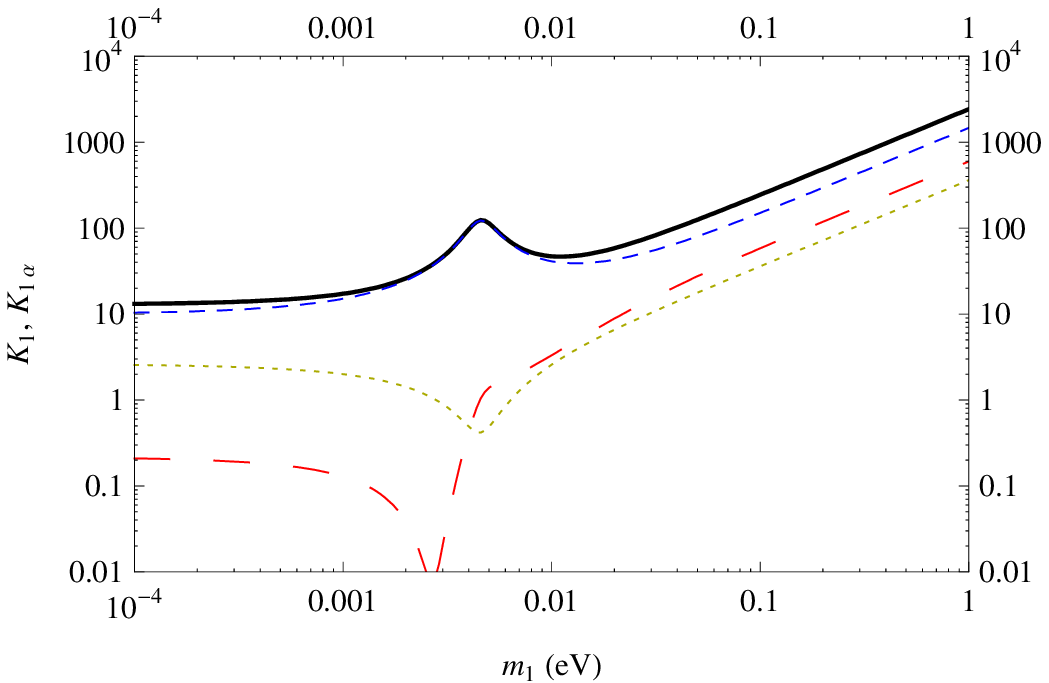,height=42mm,width=49mm}
\hspace{3mm}
\psfig{file=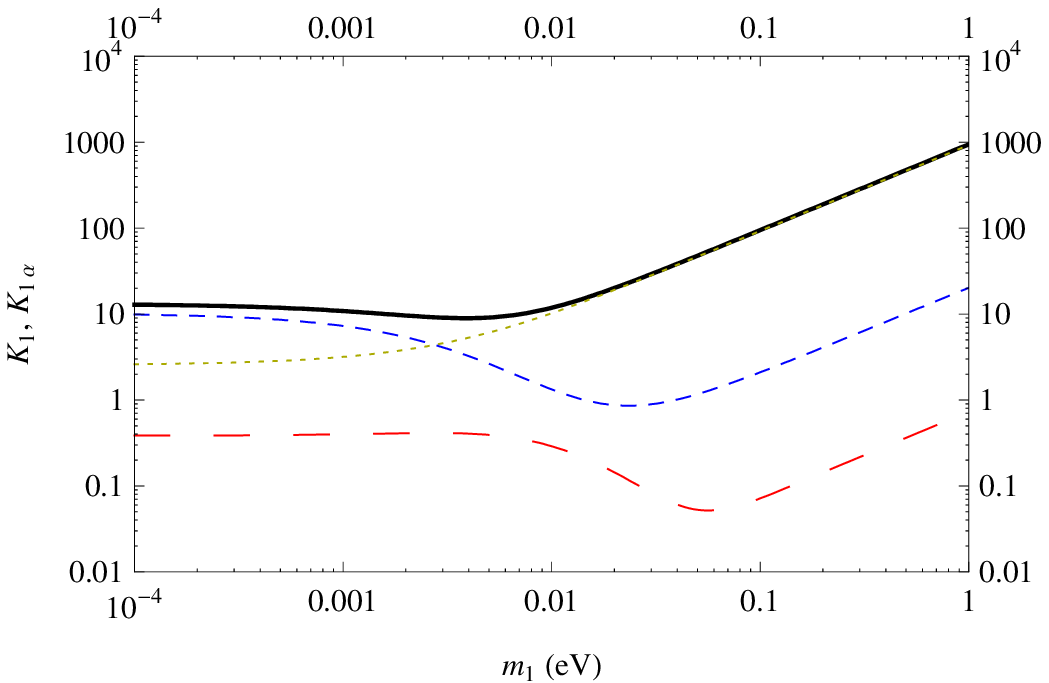,height=42mm,width=49mm}
\hspace{3mm}
\psfig{file=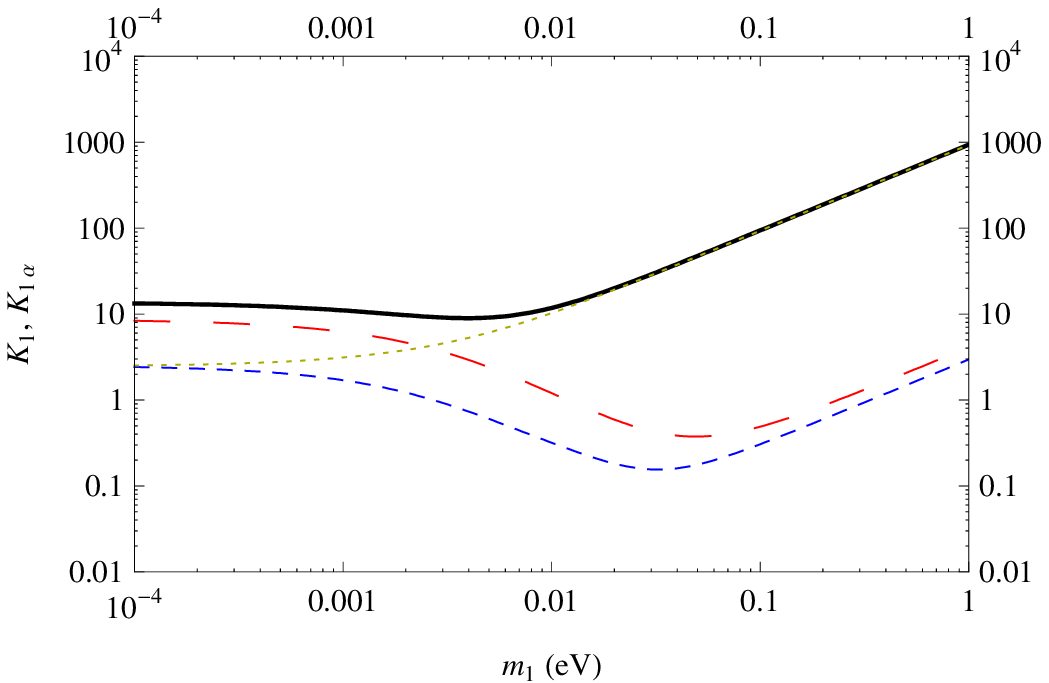,height=42mm,width=49mm}
\end{center}
\caption{Dependence of $K_1$ (black solid line), $K_{1\t}$
(red long-dashed line), $K_{1\m}$ (blue dashed line) and $K_{1e}$ (dotted
yellow line) on $m_1$ for the same value of the parameters as in Fig.~2.}
\label{fig:NBmL}
\end{figure}
The existence of regions in the space of parameters where
$K_{1\tau}\lesssim 1$, one of the key points in our discussion, can
be understood analytically. Indeed an analytical expression for
 $K_{1\tau}$ is easily found when exploiting again  the $SO(10)$ relation (\ref{SO(10)}),
\begin{eqnarray}
K_{1\tau}&=&\frac{(\alpha_3\,m_t)^2}{m_\star}\frac{\left|\left(U_R\right)_{31}\right|^2}{M_1},\,\,
|\left(U_R\right)_{31}| \simeq
\left|\frac{ \left(m_\nu\right)_{13}}{ \left(m_\nu\right)_{11}}\frac{\alpha_1 \,m_u}{\alpha_3\, m_t}\right| ,\,\,
M_1\simeq \frac{(\alpha_1\,m_u)^2}{\left| \left(m_\nu\right)_{11}\right|}\, ,\nonumber\\
%\frac{\left| \left(m_\nu\right)_{13}\right|^2}{\left| \left(m_\nu\right)_{11}\right|\,m_\star}\,,\nonumber\\
\left| \left(m_\nu\right)_{13}\right|^2&\simeq&\frac{1}{9}\left[m_1^2+m_2^2-2m_1 m_2\cos 2\rho-2s_{13}
\right. \nonumber \\
& &
\left(\sqrt{2} m_1^2\cos\left(4\rho+\delta\right)-
    \sqrt{2} m_1 m_2\cos\left(2\rho+\delta\right)+\frac{m_1 m_2}{\sqrt{2}}\cos\left(2\rho+\delta\right)\right. \nonumber \\
&-&\left.\left. \frac{m_2^2}{\sqrt{2}}
\cos\delta-\frac{3 m_1 m_3}{\sqrt{2}}\cos\left(\delta-2\sigma-2\rho\right)+\frac{3 m_2 m_3}{\sqrt{2}}\cos\left(\delta-2\sigma\right)\right)\right]\,,\nonumber\\
\left(m_\nu\right)_{11}&\simeq& \frac{2}{3}m_1 e^{-2i\rho}+\frac{1}{3}m_2\,,
\end{eqnarray}
where we have retained only terms linear in $s_{13}$. From these expressions,
we can already deduce some important results.
We see, for instance, that, for $s_{13}=0$, $K_{1\tau}$ is minimized by $\rho=0$ (mod $2\pi$)
irrespectively from the values of $\delta$ and $\sigma$, leading to
\be
K_{1\tau}^{\rm min}(s_{13}= 0)\simeq \frac{1}{3}\frac{(m_2-m_1)^2}{(2 m_1+m_2)\,m_\star}\,,
\ee
suggesting that for no mixing between the first and the third LH neutrino generation,
the wash-out of the baryon asymmetry produced by the $N_2$ decays via the interactions
mediated by the $N_1$'s
is already mild, $K_{1\tau}\sim 3$ if the LH spectrum is hierarchical,
$m_1\ll m_2\ll m_3$. The more the spectrum of the LH neutrinos is degenerate,
$m_1\simeq m_2\simeq m_3\simeq \overline{m}$,  the more the wash-out decreases, down to the value
$K_{1\tau}\simeq \left(\Delta m_{\rm sol}^2\right)^2/(36 \overline{m}^3 m_\star)\simeq 10^{-2}$. However, for $s_{13}=0$, but for non-vanishing values of
$\rho$, $K_{1\tau}$ gets very large in the degenerate case,
$K_{1\tau}\simeq (\overline{m}/3 m_{\star})(1-\cos 2\rho)/(5+\cos 2\rho)^{1/2}$.

For non-vanishing  values of $s_{13}$, one can find values of the
parameters for which $K_{1\tau}$ is smaller than unity. For instance,
for a hierarchical spectrum of LH neutrinos and independently of  the
value of  $\rho$, requiring $K_{1\tau}\lesssim 1$ leads to
\be
s_{13}\cos\left(\delta-2\sigma\right)\gtrsim \frac{m_2}{3\sqrt{2} m_3}\simeq 0.04\, .
\ee
To get the feeling of the figures involved, we may set
$\delta\simeq 2\sigma$ and  find that the wash-out
mediated by the $N_1$'s vanishes
for an experimentally allowed value of the mixing between
the first and the third generation of LH neutrinos,
$\theta_{13}\simeq 2.3^\circ$ . We stress again that these conclusions do not depend on the value of
$\alpha_2$ defined in Eq. (\ref{alpha}). These simple
analytical insights together with the numerical results
demonstrate that it is possible
to find regions of the low-energy neutrino parameters
where the wash-out mediated by the lightest RH neutrinos is totally negligible.
Furthermore, for a hierarchical spectrum of LH neutrinos,
it seems that small, but non-vanishing values
of $\theta_{13}$ may be preferred. Going back to the minimal value of $m_1$
required to have enough baryon asymmetry, we may estimate
\be
\eta_B\simeq 5\times 10^{-3}\,\ve_{2\tau}
\simeq 5\, \left(\frac{\alpha^2_2 \,
m_1}{m_3}\right)\cdot 10^{-10}\, ,
\ee
which requires
\be
m_1\gtrsim \,\left(\frac{5}{\alpha_2}\right)^2\,10^{-3}\,{\rm eV}\, ,\ee
for a normal hierarchical spectrum of light neutrinos.
%To see if these rough analytical estimates survive the requirements that  a
%large lepton asymmetry is produced by the $N_2$-decays, we now turn to a  full
%quantitative analysis.
%%%%%%%%%%%%%%%%%%%%%%%%%%%%%%%%%%%%%%%%%%%%%%%%%%%%%%%%%%%%%%%
%\section{Numerical search of the allowed regions of parameters}
%%%%%%%%%%%%%%%%%%%%%%%%%%%%%%%%%%%%%%%%%%%%%%%%%%%%%%%%%%%%%%%
In the left panel of Fig.~6
 we performed a random scan in the space of parameters,
in the case of the mixing angles within the $2\sigma$ ranges
(\ref{twosigma}) and for $\log (m_1/{\rm eV})$ spanning
in the interval $[-3.5,-0.3]$. The yellow stars are those for which
$\eta_B> 3\times 10^{-10}$ while the green circles, found with a frequency
$\sim 10^{-4}$, are those corresponding to successful leptogenesis
for $\eta_B > 5.9 \times 10^{-10}$.
\begin{figure}
\begin{center}
\psfig{file=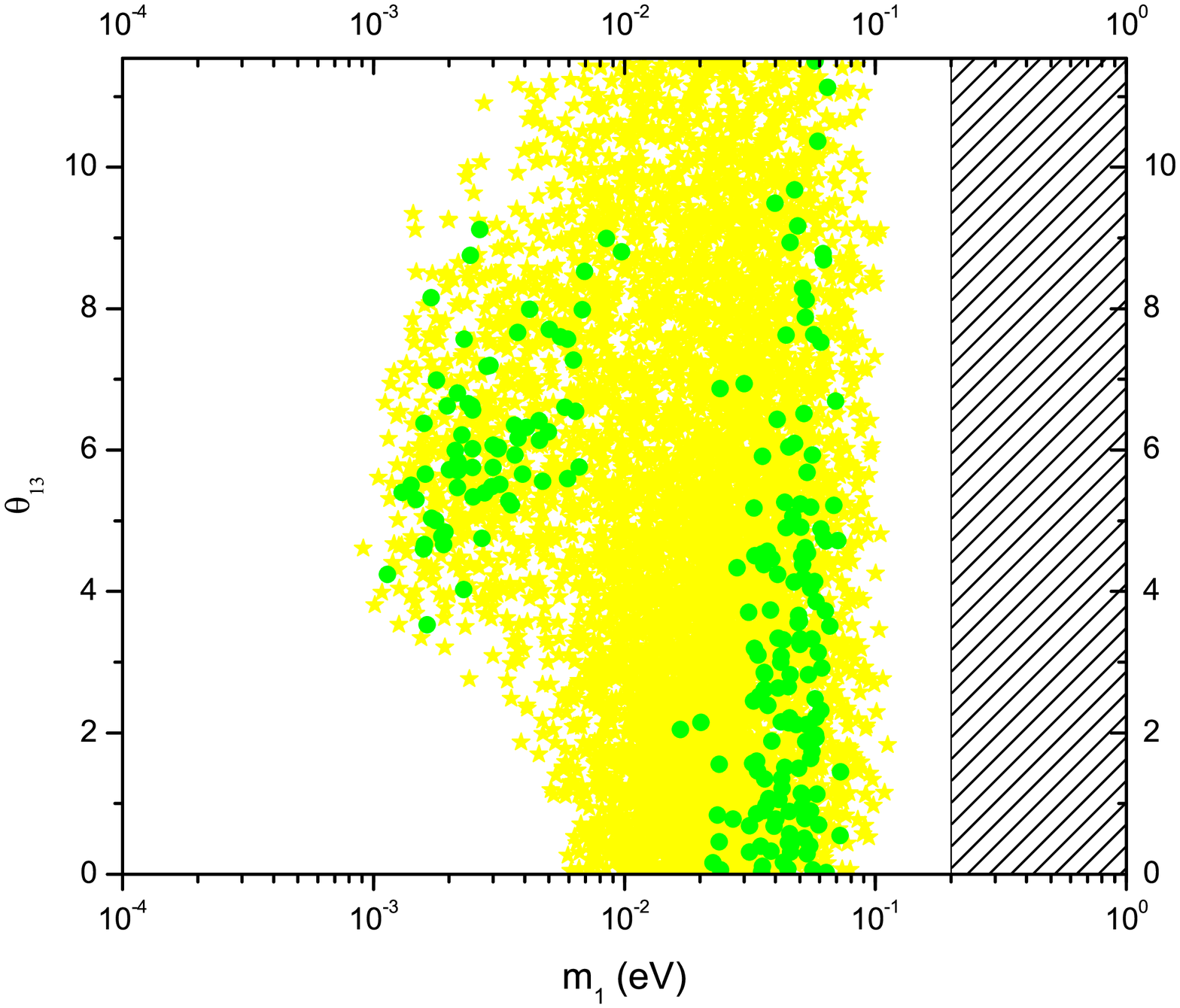,height=63mm,width=75mm}
\hspace{5mm}
\psfig{file=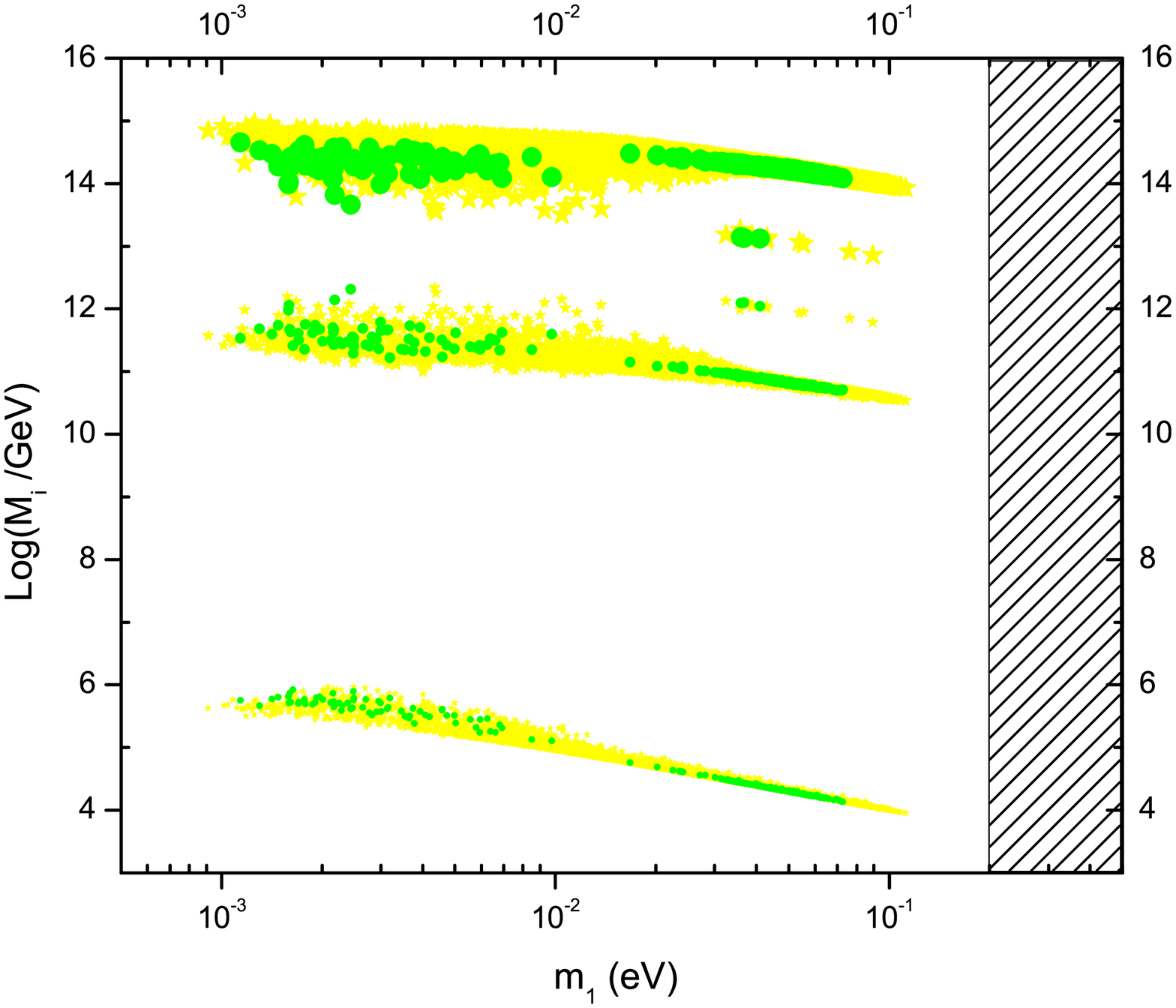,height=63mm,width=75mm}
\caption{
Scan plot in the $m_1-\theta_{13}$ plane (left panel) for $V_L=I$ and $\alpha_i={1,5,1}$.
The hatched area is excluded by the cosmological upper bound on $m_1$ (cf. (\ref{upperbound})).
Yellow stars and green circles correspond respectively to
$\eta_B> 3\times 10^{-10}$ and $\eta_B > 5.9\times 10^{-10}$
and corresponding values of the RH neutrino masses $M_i$ (right panel).
Notice that all points have been obtained imposing $M_3/M_2>10$ and therefore a
resonant $C\!P$ asymmetry enhancement is not necessary for the mechanism to work.}
\label{fig:vanilla}
\end{center}
\end{figure}
We imposed $M_3/M_2>10$ and therefore, we can conclude that
successful leptogenesis is possible without relying on any
$C\!P$ asymmetry enhancement due to $M_2\simeq M_3$. In the right panel of
Fig.~6 we also show the corresponding values of the three RH neutrino masses in the
 $M_i-m_1$ plane. We see that
successful leptogenesis may be achieved for values of $M_2$ as small as
  $10^{10.5}$ GeV. On the other hand one can see that there is a lower
bound $m_1\gtrsim 10^{-3}\,{\rm eV}$ and that for a fully hierarchical
spectrum ($m_1\lesssim m_{\rm sol}\simeq 10^{-2}\,{\rm eV}$)
only non-vanishing values of $\theta_{13}$ in the range
$\sim (3^{\circ}-9^{\circ})$ are allowed.
A more detailed analysis of the allowed values
of all parameters will be presented in \cite{inprep}.

%%%%%%%%%%%%%%%%%%%%%
\section{Conclusions}
%%%%%%%%%%%%%%%%%%%%%
In this paper we have studied the thermal leptogenesis mechanism
within $SO(10)$-inspired models where the LH neutrino masses are generated
via the type I see-saw mechanism and $SO(10)$ inspired conditions are
imposed on the neutrino Dirac mass matrix. We have shown that a large  enough
baryon asymmetry may be created  by the decays of the next-to-lightest
RH neutrinos when flavor effects are accounted for. Indeed, it is possible
to find regions of the parameter space where the asymmetry in a given flavor,
in our case the tau one, is not erased by the processes mediated at much lower
temperatures by the lightest RH neutrinos. This result has been obtained
assuming a hierarchical spectrum of RH neutrinos, therefore without resorting
to any resonant enhancement factor in the $C\!P$ asymmetries.  Let us also  remark that
for thermal leptogenesis to be valid, the initial temperature for the evolution of the
system has to be higher than about $M_2/5$ and from the right panel of Fig.~6 one can see
that values of the reheating temperature as small as $10^{10}\,{\rm GeV}$
may be sufficient for generating a large enough baryon asymmetry.
This point becomes relevant when  considering the super-symmetrized
version of the scenario where the gravitino bound on the reheating temperature
after inflation \cite{grav} becomes an issue.
We have also found both analytically and
numerically an interesting lower bound on the mass of the lightest LH neutrino,
$m_1\gtrsim 10^{-3}\,{\rm eV}$, and that non-vanishing values of the
mixing angle $\theta_{13}$ are preferred.
A more complete analytical and numerical analysis, relaxing the assumptions made here for
simplicity ({\it e.g.} $V_L=1$, normal scheme for LH neutrinos),
is now in preparation \cite{inprep}.

When this paper was being prepared, Ref. \cite{french} appeared where
flavored leptogenesis based on a mixed type I plus type II
see-saw mechanism was studied. The comparison between our results and those in Ref.
\cite{french} is made difficult by the fact that there the type I see-saw case has been marginally studied as a limit of a more general left-right framework. We believe we have found regions of the parameter space
where leptogenesis is successful, not fully investigated in Ref. \cite{french}.
 In particular, in our study the asymmetry is dominantly produced along the tau flavor
 and we have given more emphasis to the impact of low energy parameters.

\vskip 0.5cm
\centerline{\bf Acknowledgements}

\noindent
We wish to thank Gian Giudice and Ferruccio Feruglio for useful discussions on deviations
from the exact $SO(10)$ mass relations. We also thank Asmaa Abada for useful
correspondence on Ref. \cite{french}.
This research was supported in part by the European Community's Research
Training Networks under contracts MRTN-CT-2004-503369 and
MRTN-CT-2006-035505. PDB is supported by the Helmholtz Association of
National Research Centres, under project VH-NG-006.

\end{document}